\documentclass[aps,prl,preprint,groupedaddress]{revtex4-2}
\usepackage{graphicx}
\usepackage{dcolumn}
\usepackage{bm}
\usepackage{mathtools}
\usepackage[utf8]{inputenc}
\usepackage{extarrows}
\usepackage[T1]{fontenc}
\usepackage{booktabs, array, mathptmx, float, tabularx, booktabs, lipsum, amsmath,multirow}
\usepackage{siunitx, xcolor}
\usepackage[version=4]{mhchem}
\graphicspath{{figs/}{figsgaoerb/}} 
\usepackage[colorlinks,linkcolor=blue,anchorcolor=blue,citecolor=blue]{hyperref}
\begin{document}
\title{Bounced Model of Droplet on Moving Substrate}
\author{Chengwu Liu}
\email{202023102003@mail.sdu.edu.cn}
\homepage{https://orcid.org/0000-0001-9067-1892}
\altaffiliation{School of Physics, Shandong University, Jinan 250100, China.}

\affiliation{School of Physics, Shandong University, Jinan 250100, China.}
	\begin{abstract}
	Firstly, we get the completely bouncing criteria Cr for droplet on moving substrate. The bouncing without splashing condition is $	\mathrm{Cr}>1$. Then, we mainly research the effect of wind field for droplet, and get the completely bouncing criteria $\mathrm{Cr}_{\mathrm{wind}}$ for droplet with wind. 
	
	Lastly, we get the contact angle of droplet on the moving substrate and calculate the Time Independent Reynolds Equation with $rho$ and $\mu$ are constant. 
	\end{abstract}
\maketitle
		\section{Introduction}
The questions of droplet on a surface are related to the interaction of interface. 

There is a micrometer-size gas film in the interface between liquid and solid. This gas film was firstly observed by the way of snapshoot \cite{1}. The evolutionary process of gas film was firstly observed by X-Ray technology \cite{2} at the moment of contacting. They found that the gas film evolve to a bubble with spending on microsecond-size time. E. Sawaguchi \cite{3} found that the distribution of thickness of droplet on a moving surface is similar to saddle surface. In addition, the hydrophobicity of droplet on a moving surface is enhanced and is similar to Leidenfrost effect \cite{4}. Therefore, the interaction between liquid and solid would be affected by the motion of surface. In this paper, we will talk about how these parameters affect the hydrophobicity in section 2.

Ted Mao \cite{5} assumed a critical bounced state to deal with the question of bounced on motionless surface and got a critical bounced criteria $E_{\mathrm{ERE}^*}$. Actually, the gas film on motionless surface is different from gas film on moving surface. So the interaction of liquid and solid on motionless surface is also a little different from one on moving surface. In this paper, we will talk about this question in section 3.

Droplet also might splash on a solid surface. We have some models to describe splashing \cite{6}\cite{7}\cite{8}\cite{9}\cite{10}. In this paper, we will talk about how extra wind affects the splashing and bounced of droplet in section 4. 
	\section{Droplet on the moving substrate}
	The final states of droplet after it impacts on moving substrate are various. We name bouncing without splashing as completely bouncing, bouncing and splashing as partially bouncing. We find out many physical phenomenons about droplet on the moving substrate with lots of experiments(The experiments conditions are in the supplemental material\cite{SM}. ). For instance, there is the critical speed which is the shift of bouncing and retention. We will deeply research the completely bouncing condition and critical speed below.
	\subsection{Completely Bouncing Criteria for droplet on Moving Substrate}
	We will research the phenomenons without splashing in this part. 
	
	Droplet will spread, retract, bounce/retain after impacting on the moving substrate. According to the different interaction modes on the solid-liquid interface, the interface can be divided into three areas. As the figure 2(a) shown, the first one is gas film(thickness $h\sim 10\mu\mathrm{m}$). The cohesive force of gas film acts as normal capillary force to inhibit bouncing. The second one is solid-liquid interface. The interaction of liquid and solid is Van der Waals force. The third one is molecule area. The interaction of this part mainly include the Derjaguin Disjoining Pressure $\Pi\left( h\right) =\Pi_{\mathrm{vdW}}+\Pi_{\mathrm{EL}}+\Pi_{\mathrm{struc.}}+\Pi_{\mathrm{steric}}$. In these areas, inhibition and promotion effects for bouncing should be researched in detial. 
	\subsubsection{Inhibition Effects}
	Firstly, the maximal viscous dissipation of droplet could be approxed is $D\le V_G=mgh\sim 10^{-3}\mathrm{J}$ with enery conservation(we order the surface of substrate is the zero gravitational potential surface.) Nextly, the dissipation of droplet also due to the microstructure on the surface of substrate. The roughness could be evaluated by the size of roll angle in general. We could ignore the dissipation of this part with considering a small roll angle(smooth surface).  Acturally, we use the substrate with small roll angle in our experiments(more detials in supplemental material\cite{SM}). Then, we will consider the interaction between liquid and solid on the gas film and solid-liquid interface. The attraction is Van der Waals force mainly. And it is the potential force and plays an obvious role on micrometer scale justly. Therefore,  the vdW force between solid and liquid has no effect on bouncing. However, some gas are carried in the gas film during the droplet falling process. This gas film will go away from the system constituted by droplet and solid substrate during the bouncing process. And the capillary force of gas film will dissipate the energy of droplet. So, the cohesive force of gas film will inhibit bouncing. Nextly, we will explain it detially. 
	
	First of all, this process is different from the previous section. In this section, the pressure of gas film $p$ changes continuously over time, and it's relatively large. The kinetic equation could be given through analysing the droplet as the figure 2(b) shown.
	\begin{equation}
		\left( p-p_\ominus\right) S=m\frac{\mathrm{d^2}H}{\mathrm{d}t^2}+mg
	\end{equation}
	where $p_\ominus$ is atmospheric pressure, $H$ is the vertical positionsl coordinaten of droplet, $S$ is the contact area of liquid and solid. Apparently, we could find out $H$ and $S$ change over time through observing experiments results. So $p$ also should change over time as figure 2 shown. Then the dissipation due to the cohesive force of gas film is be considered below. 
	
	The gas film is divided into two parts $g_1,\:g_2$ with $h$ thickness as figure 2(d) shown. And two parts at a distance of $d\sim 10^{-9}\mathrm{m}$($d$ is the distance of two molecule). Hence, the per unit area vdW potential between $g_1$ and $g_2$ is 
	\begin{equation}
		w_g =-\frac{A}{12\pi}\left[ \frac{1}{d^2}+\frac{1}{\left( d+2h\right) ^2}-\frac{2}{\left( d+h\right) ^2}\right]\xlongequal{h\gg d}-\frac{A}{12\pi d^2}
	\end{equation}
	where $A\sim 10^{-19}\mathrm{J}$ is Hamker constant. We assume that the thickness of gas film changes from $h_0$ to $h$ from the state of maximum spreading to bouncing exactly, $\tau'$ is the time interval of this process. Then the energy dissipation during retraction process is
	\begin{equation}
		Q_{\mathrm{cap}}=\left| \int_{h_0}^{h}-\frac{\partial w_g}{\partial d}S\mathrm{d}h\right| =\left| \int_{0}^{\tau'}\frac{AS}{6\pi d^3}\frac{\mathrm{d}h}{\mathrm{d}t}\mathrm{d}t \right|
	\end{equation}	
	Then we expand $h$ to linear term and consider the ideal gas hypothesis. 
	\begin{equation}\label{taylor expansion and ideal gas hypothesis}
		\frac{\mathrm{d}h}{\mathrm{d}t}=\frac{h-h_0}{t},\:\:d^3=\frac{kT}{p}
	\end{equation}
	We could give the energy dissipation due to capillary force with equation \ref{taylor expansion and ideal gas hypothesis}.
	\begin{equation}\label{energy dissipation due to capillary force}
		\begin{split}
			Q_{\mathrm{cap}}&\approx\left| \int_{0}^{\tau'}-\frac{ApS\left( h-h_0\right) }{6\pi d^3t}S\mathrm{d}t\right| \\
			&\xlongequal[Sh_0=V_0]{Sh=V}\left| \int_{0}^{\tau'}\frac{Ap\left( V-V_0\right) }{6\pi kTt} \mathrm{d}t\right|\\
			&\xlongequal[p_0V_0=\nu RT]{pV=\nu RT}\frac{A\nu R}{6\pi k}\left| \int_{0}^{\tau'}\frac{1-\left( p/p_0\right) }{t}\mathrm{d}t\right| 
		\end{split}
	\end{equation}
	This is a improper integral. In order for the improper integral is convergent. We have
	\begin{equation}
		\lim_{t\to 0+}\frac{1-\left( p/p_0\right) }{t}=\lim_{t\to 0+}-\frac{1}{p_0}\frac{\mathrm{d}p}{\mathrm{d}t}=a\Rightarrow \frac{\mathrm{d}p}{\mathrm{d}t}=-ap_0
	\end{equation}
	Therefore, we have the pressure $p$ changes linearly with time below the above approximation. Then bring equation \ref{taylor expansion and ideal gas hypothesis} to equation \ref{energy dissipation due to capillary force}.
	\begin{equation}
		\begin{split}
			Q_{\mathrm{cap}}&=\frac{A\nu R}{6\pi k}\left| \int_{p_0}^{p^\ominus }\frac{\mathrm{d}p}{p_0}\right|=\frac{A\nu N_{\mathrm{A}}}{6\pi }\frac{p_0-p^{\ominus}}{p_0}=\frac{A\nu N_\mathrm{A}}{6\pi }a\tau'\sim 10^{-3}\mathrm{J}
		\end{split}
	\end{equation}
	where $\nu$ is the amount of substrate of gas film, $N_{\mathrm{A}}$ is Avogadro constant. We could conclude that the order of energy dissipation due to capillary force is same as the gravitational potential energy. So it couldn't be ignored.
	
	Lastly, let's consider the viscous dissipation. It's so difficult to calculate that we have to estimate it. Let's consider an  undemanding toy model. 
	
	Firstly, the viscous dissipation should relateto spreading factor with a certain function. The viscous dissipation which droplet impacts on a smooth motionless substrate relate to $\left( d_m/D\right)^{2} $\cite{16} during spreading process, and relate to $\left( d_m/D\right)^{2.3} $\cite{5} during retraction process. Hence, we associate the spring oscillator model of two degrees of freedom with this question(Two springs placed perpendicular to each other horizontally on a smooth substrate). We assume that the viscous dissipation is $E_{\mathrm{diss}}\sim QE_{\mathrm{p\:max}}$, where $E_{\mathrm{p\:max}}$ is the maximum spreading "elastic potential energy" for the first time, $Q$ is similar to quality factor.
	\begin{equation}
		E_{\mathrm{diss}}\sim \alpha k_n \left( D_{\mathrm{n\:max}}-D_0\right)^2 +\beta k_t \left( D_{\mathrm{t\:max}}-D_0\right) ^2
	\end{equation}
	Considering that the effect of tangential mainly due to surface tension and viscous shear stress $T_{\nu}$, the effect of normal mainly due to surface tension. We could give the $k_n$ and $k_t$ using above model. 
	\begin{alignat}{2}
		k_\mathrm{t}&\sim \frac{a\gamma D_\mathrm{t\:max}+bT_\nu}{D_{\mathrm{t\:max}}-D_0} \\
		k_\mathrm{n}&\sim \frac{a' \gamma D_\mathrm{n\:max}}{D_{\mathrm{n\:max}}-D_0} 
	\end{alignat}
	where $\alpha,\:\beta ,\:a,\:b,\:a'$ is constant, $T_\nu\sim\eta VD_{\mathrm{n\:max}}D_{\mathrm{t\:max}}/\delta$ is the viscous shear stress\cite{14}, $\delta$ is the thickness of gas film estimated by LLD's law\cite{12}, $V$ is the speed of substrate, the maximum tangential spreading diameter on moving substrate\cite{12} and the maximum spreading diameter on motionless substrate\cite{15} are respectively $D_\mathrm{t\:max}/D_0\sim \mathrm{We}^{1/4}\mathrm{Ca}^{1/6}$ and $D_\mathrm{max}/D_0\sim \mathrm{We}^{1/4}$. And the substrate speed has no effect on normal maximum spreading diameter. Therefore, $D_\mathrm{n\: max}\sim \mathrm{We}^{1/4}$. Then, we could estimate the viscous dissipation with scaling law. 
	\begin{figure}[!htbp]
		\centering
		\label{figure2}
		\includegraphics[width=8.6cm]{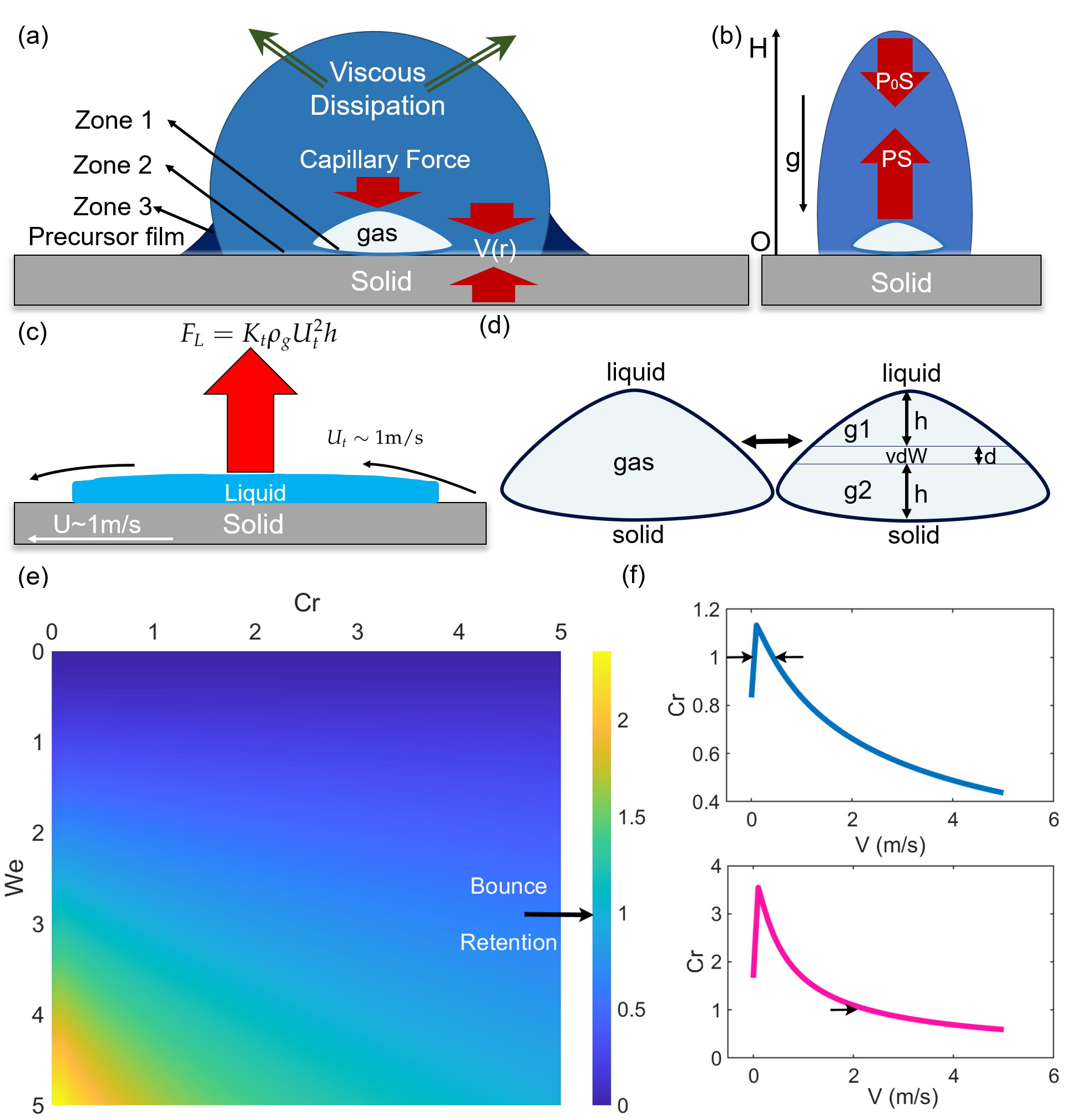}
		\caption{(a)The interface can be divided into three areas. The first one is gas film(Zone 1). The second one is solid-liquid interface(Zone 2). The third one is precursor film(Zone 3). The interaction of this part mainly include the Derjaguin Disjoining Pressure. In addition to these, energy dissipation(viscous dissipation and energy dissipation due to capillary force) also play an important role on bouncing. (b)Pressure isn't a constant during the retraction process. (c)Effect of wind field caused by moving substrate on droplet. This lift force could be ignored when the speed of substrate is little.(d)The gas film is divided into two parts $g_1,\:g_2$ with $h$ thickness. (e)Completely Bouncing Criteria for droplet on Moving Substrate $\mathrm{Cr}$. The droplet will retain on the substrate if $\mathrm{Cr}\le1$, and bounce on the substrate if $\mathrm{Cr}>1$. (f)Completely Bouncing Criteria for droplet on Moving Substrate $\mathrm{Cr_{V}}$, which is only due to the speed of substrate. The arrow points to the critical speed. }
	\end{figure}
	\subsubsection{promotion effects}
	Firstly, one of the promotion effects is initial kinetic energy $E_{\mathrm{total}}=\frac{1}{2}\rho\left( \frac{1}{6}\pi D_0^3 \right) U^2$, where $U$ is the speed which droplet impacts the moving substrate. Next, as the figure 2(c) shown. The movment of solid substrate caused the flow of air beacuse air is the viscous fluid. Hence, the pressure around substrate will be decreased by wind field. So, it has a lift force $F_L\sim \rho_gU_t^2h$ on droplet, where $\rho_g=1.185\mathrm{kg/m^3}$ is the density of air(25℃), $\rho=997\mathrm{kg/m^3}$ is the density of water(25℃), $U_t\sim 1\mathrm{m/s}$ is wind speed around the substrate, $h\sim 10^{-3}\mathrm{m}$ is the maximum spreading thickness of droplet, $D\sim 10^{-3}\mathrm{m}$ is the initial diameter of droplet. So, we have 
	\begin{equation}
		\frac{\rho DU^2}{\rho_gU_t^2h}\sim 10^{3}
	\end{equation}
	So lift force could be ignored when the speed of substrate is little($U\sim 1\mathrm{m/s}$). The situation of big wind speed will be discussed in section 4. 
	\subsubsection{Completely Bouncing Criteria for droplet on Moving Substrate}
	Hence, we could conclude that the initial kinetic energy $E_{\mathrm{total}}$ promote to bounce, viscous dissipation $E_{\mathrm{diss}}$ and energy dissipation due to capillary force $Q_{\mathrm{cap}}$ inhibit bouncing. 
	
	Considering a imaginary state which droplet bounce exactly. Then, we could get the condition of bouncing using energy conservarion. We have 
	\begin{equation}
		E_{\mathrm{total}}+mg\frac{D}{2}>E_{\mathrm{Diss}}+Q_{\mathrm{cap}}+mg\frac{D}{2}
	\end{equation}
	So, we can get the completely bouncing criteria for droplet on moving substrate. 
	\begin{equation}\label{Completely Bouncing Criteria for droplet on Moving Substrate}
		\mathrm{Cr}=\frac{\pi \mathrm{We}^{3/2}\sqrt{\gamma/D_0\rho}}{\frac{2A\nu N_{\mathrm{A}}}{\pi D_0^2\gamma}\left ( \frac{p_0-p^\ominus }{p_0} \right ) +12\left [ \left ( \beta\mathrm{We}^{1/4}\mathrm{Ca}^{1/6}+\frac{\beta'D_0\mathrm{We}^{1/2}\mathrm{Ca}^{1/2}}{l\mathrm{c}} \right ) \left ( \mathrm{We}^{1/4}\mathrm{Ca}^{1/6}-1\right ) +\alpha \mathrm{We}^{1/4}\left ( \mathrm{We}^{1/4}-1 \right )  \right ] }
	\end{equation}
	where $l_c=\left( \gamma/\rho g\right) ^{-1/2}$ is the Capillary length of water, $rho$ is the density of liquid, $\gamma$ is the liquid-gas surface tension, $\alpha,\:\beta,\:\beta'$ are the constants to be determined. The numerator is shown that the effect of initial kinetic energy. The left term of denominator is shown that the effects of capillary force in the gas film. The right term of denominator is shown that the effects of viscous dissipation. The bouncing without splashing condition is 
	\begin{equation}
		\mathrm{Cr}>1
	\end{equation}
	And completely bouncing criteria for droplet on moving substrate $\mathrm{Cr_{V}}$, which is only due to the speed of substrate. 
	\begin{equation}\label{speed criteria}
		\mathrm{Cr_{V}}=\frac{\mathrm{A}}{\mathrm{B}+\left( \mathrm{C}V^{1/6}+\mathrm{C'}V^{1/2}\right) \left( \mathrm{D}V^{1/6}-1\right) }
	\end{equation}
	where A,B,C,C',D are indepent to substrate speed $V$. The droplet will retain on the substrate if $\mathrm{Cr_{V}}\le 1$, and bounce on the substrate if $\mathrm{Cr_{V}}>1$. 
	
	In a word, the completely bouncing criteria for droplet on moving substrate is relate to capillary number $\mathrm{Ca}$ and Weber number $\mathrm{We}$ which both are the initial state parameters of the droplet and moving substrate, as figure 2(e) shown. And we also explain the experimental phenomenons which the final state of droplet will shift with changing substrate speed as figure 2(f) shown. We also could find out the critical speed which is the shift of bouncing and retention as figure 2(f) shown. These results show that droplet might undergo the transformation of "Retention to Bounced" or "Retention to Bounced to Retention". But we didn't consider the dissipation due to the microstructure on the surface of substrate. It's also a complex question
	\section{The Effect of Wind Field for Droplet}
	In the discussion of previous section, we find that a lift force due to the wind field couldn't be ignored with high substrate speed. The lift force promote to bounce. We even find out the splashingofdroplet in further experiments. So we design the experiment(more detials are in supplemental material\cite{SM}) to illustrate the importance of wind field for the final state of droplet as figure 3(a)(b)shown. 
	\begin{figure}[!htbp]
		\centering
		\label{figure3}
		\includegraphics[width=8.6cm]{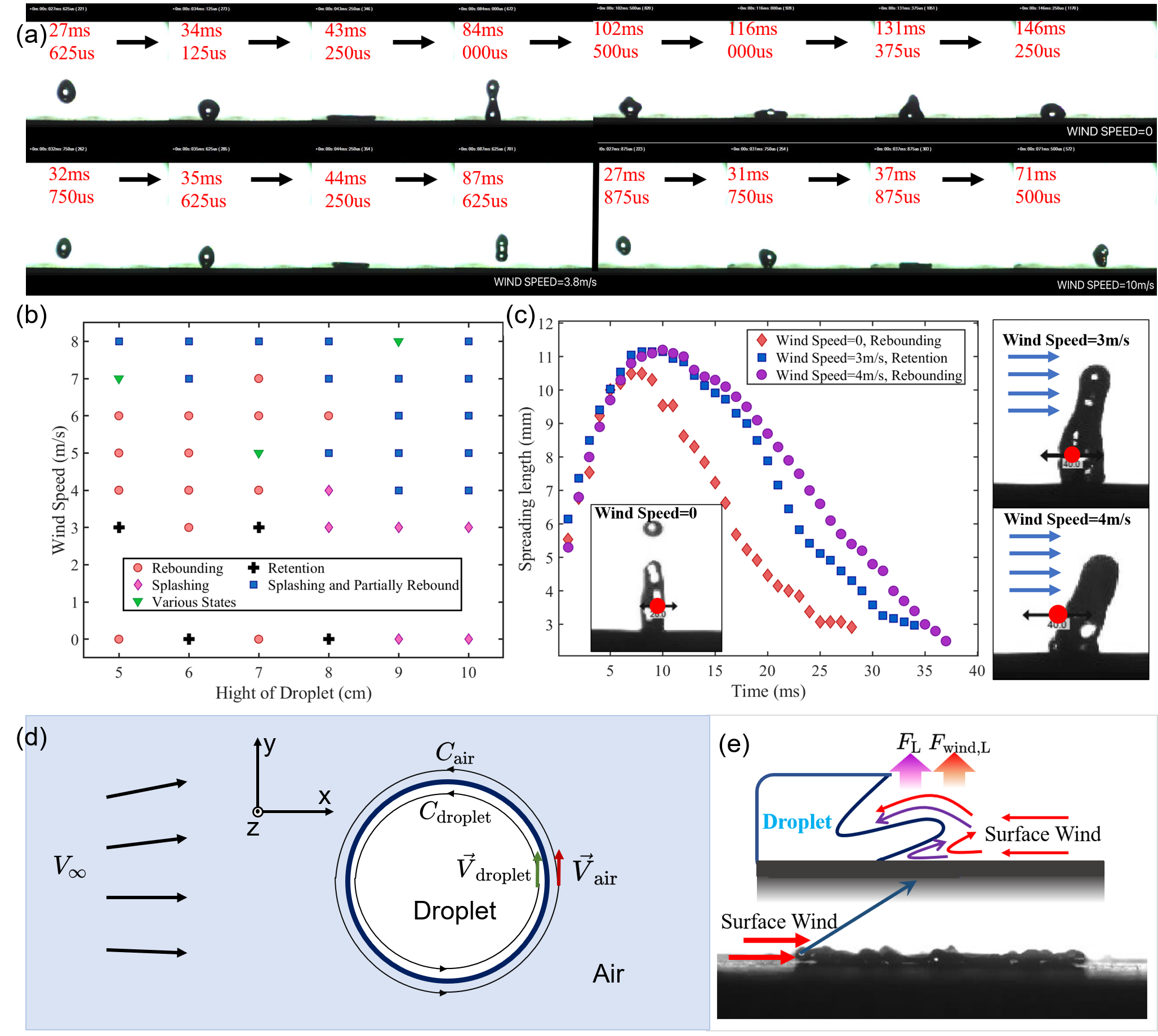}
		\caption{(a)The arrow represents the direction in which time is passing. It shows the droplet state with wind speed is 0, 3.8m/s and 10m/s. (b)The finial state figure of dropplet. Three points which have various states respectively represent that splashing+partially bouncing and completely bouncing(coordinates are (7,5) and (9,8)), various states(coordinate is (5,7)). (c)The spreading of droplet which inital height is 5cm in wind field. The red points are the initial position of the center of mass(d)Considering the loops $C_{\mathrm{droplet}}$ and $C_{\mathrm{air}}$. (e)R\&G model with wind field. $F_{\mathrm{wind,\:L}}$ is the extra effect on liquid finger. }
	\end{figure}
	We conclude that the final state of droplet is affected by both the initial height of droplet(We) and the speed of wind(Ca) from figure 3(b). And the state of droplet will shift to splashing and partially bouncing with creasing the wind speed and the initial height. The state of droplet will shift to splashing with only creasing the initial height. The state of droplet with small initial height will shift to splashing and partially bouncing if only creasing the wind speed. And state also could shift from retention to completely bouncing/from completely bouncing to retention. We conclude that the spreading factor et. al. could change with the wind speed from figure 3(c). Next, we will research the causes of these phenomenons.
	\subsection{The Transition Between Retention and Completely Bouncing}
	Firstly, we assume that (1)all flows could be ragard as the isentropic flows. (2)The boundary conditions between liquid and gas obey Navier boundary conditions, i.e.,  $\vec{V}_{\mathrm{droplet}}=\vec{V}_{\mathrm{air}}$(3)Wind field is 2D incompressible laminar flow, i.e., $\nabla\cdot \vec{V}=0,\:V_\mathrm{z}=\mathrm{const.}$.
	
	We take two circuits $C_{\mathrm{droplet}}$ and $C_{\mathrm{air}}$ around the interface between liquid and gas. The velocity circulations respectively are
	\begin{equation}
		\Gamma_{\mathrm{droplet}}=\oint\limits_{C_{\mathrm{droplet}}}\vec{V}_{\mathrm{droplet}}\cdot\mathrm{d}\vec{l}=\Gamma_{\mathrm{air}}=\oint\limits_{C_{\mathrm{air}}}\vec{V}_{\mathrm{air}}\cdot \mathrm{d}\vec{l}=\Gamma
	\end{equation}
	These velocity circulations $\Gamma$ are constant because of assumation 2. And the interaction of wind field on the droplet could be divided into vertical and horizontal direction. The bouncing state mainly depends on the vertical interaction. 
	\begin{equation}
		\vec{L}=\rho_a\vec{V}_{\infty}\times\vec{\Gamma}_{\mathrm{air}}=-\rho_a\Gamma\vec{V}_{\infty}\times \vec{k}
	\end{equation}
	It can be seen from assumation 3
	\begin{equation}
		\nabla\times \vec{L}=-\rho_a\Gamma\left[ \left( \vec{k}\cdot \nabla\right)\vec{V}_{\infty}+\left( \nabla\cdot\vec{k}\right) \vec{V}_{\infty}-\left( \vec{V}_{\infty}\cdot{\nabla}\right) \vec{k}-\left( \nabla\cdot \vec{V}_{\infty}\right) \vec{k} \right] =0
	\end{equation}
	So the lift force $\vec{L}$ is potential force. We assmue the initial height of droplet is $h_0$ and the lift force potential of initial position is 0. Then the lift force potential is
	\begin{equation}
		V_{\mathrm{L}}\left( y\right) =-\int -\rho_a\Gamma\vec{V}_{\infty}\times \vec{k}\cdot\vec{j}\mathrm{d}y=-\rho_a\Gamma\int V_{\infty\mathrm{x}}\mathrm{d}y=\rho_a\Gamma\int_{h_0}^{y}V_{\infty\mathrm{x}}\mathrm{d}y
	\end{equation}
	So the completely bouncing criteria for droplet is 
	\begin{equation}
		\mathrm{Cr_{wind}}=\frac{\rho g\pi D_0^3\left( h_0-0.5D_0\right) /6}{E_{\mathrm{Diss}}+Q_{\mathrm{cap}}+\rho_a\Gamma\int_{h_0}^{D_0/2}V_{\infty\mathrm{x}}\mathrm{d}y}
	\end{equation}
	The bouncing without splashing condition is 
	\begin{equation}
		\mathrm{Cr_{wind}}>1
	\end{equation}
	The final state of droplt could shift between bouncing and retention beacuse $\Gamma\int_{h_0}^{D_0/2}V_{\infty\mathrm{x}}\mathrm{d}y$ could bigger/smaller than 0.
	\subsection{The Transition Between Splashing and without Splashing}
	Then, considering the transition between splashing and without splashing. The front of droplet maybe generate the liquid finger during the spreading process. Liquid finger will be not only affected by the lubrication force of bottom gas and the attraction of top gas\cite{9}$F_{L}=K_l\nu_gV_t+K_u\rho_gV_t^2H_t$, but also affected by the wind field as figure 3(e) shown. The effect of extra wind field is shown as lift force $F_{\mathrm{wind,L}}\left( V\right) $, where $V$ is the speed of wind speed. Hence, we could introduce the $F_{\mathrm{wind,L}}\left( V\right) $ to R\&G model.
	\begin{equation}
		\beta^2_{\mathrm{Wind}}=\frac{F_{\mathrm{L}}+F_{\mathrm{wind,L}}}{2\gamma}
	\end{equation}
	So, the state of droplet maybe have the transition between splashing and without splashing with the wind field.

	\section{The Balance of Droplet on the Moving Substrate}
	\subsection{The Contact Angle of Droplet}
	The bottom of droplet will generate a very thin gas film when it impact on a substrate\cite{2}. The gas film will be saddle shape stably on moving substrate\cite{3}. The pressure of gas film is $p\sim 10\mathrm{Pa}$\cite{3},  the gas film thickness is $\delta\sim 10\mathrm{\mu m}$. We also could approx the Knudsen number of gas film is $K_n=\frac{\lambda}{h}\sim 10^{-4}$. So the gas film could be regard as the continuons flow. If we assume that pressure is a constance on the direction perpendicular to moving substrate, the thickness and pressure of lubrication gas obey the Reylond Equation: 
	\begin{equation}
		\frac{\partial}{\partial x}(\frac{h^3}{\mu}\frac{\partial p}{\partial x})+\frac{\partial}{\partial y}(\frac{h^3}{\mu}\frac{\partial p}{\partial y})=6U\frac{\partial h}{\partial x}
	\end{equation}	
	where $U$ is the speed of moving substrate, $\mu$ is the dynamic viscosity of lubrication gas. This equation shows that the relation of distribution between thickness and pressure. We could give the distribution of pressure with measuring the distribution of thickness. 
	\begin{figure}[!htbp]
		\centering
		\includegraphics[width=8.6cm]{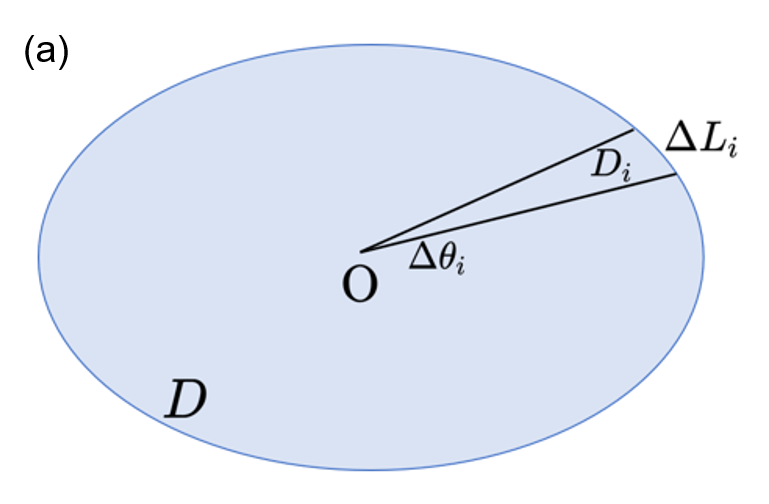}
		\caption{The diagrammatic map of gas film.}
	\end{figure}
	Then we research the effect of contact angle on moving surface. Considered the difference form gas film on motionless and moving substrate originates from the moving substrate. So, we focus on this element, then analyse this question with minimum energy principle. 
	
	We order that the area of gas film is D and O is the center of area D, as shown in the figure1, we research the infinitesimal area $\mathrm{D_i}$ with infinitesimal angle and radial length $R_i$. It is full of air and saturated vapour in area $\mathrm{D_i}$, the pressure of vapour obeys the Clapeyron Equation $p=p_0\exp(-L_{\mathrm{v,m}}/RT)$. And the molecular number and pressure of two components(saturated vapour and air) obey that 
	\begin{equation}
		1=\frac{N_{\mathrm{air}}}{N}+\frac{N_{\mathrm{H_2O}}}{N}
		,\:1=\frac{p_{\mathrm{air}}}{p}+\frac{p_{\mathrm{H_2O}}}{p}
	\end{equation}
	The mean kinetic energy of two components can be given that $\bar{e}_{\mathrm{air}}=\frac{5}{2}kT,\:\bar{e}_{\mathrm{H_2O}}=3kT$, if air is regarded as diatomic molecule. So we could give the internal energy of gas film
	\begin{equation}\label{internal energy}
		E_k=\sum_{i=1}^{n}\iint\limits_{D_i}nh\bar{e}\mathrm{d}\sigma=\sum_{i=1}^{n}\iint\limits_{D_i}\frac{5}{2}ph+h\frac{p_0}{2}\exp(-\frac{L_{v,m}}{RT})\mathrm{d}\sigma
	\end{equation}
	Hence, assumed that the front of droplet has a infinitesimal virtual displacement $\delta R_i$. Approximately, the interfacial energy between solid and gas remain unchanged because of the gas film in the solid-gas interface. So the variation in energy of system is $\delta E_i=(\Delta L_i\cos\theta_L \gamma_{LG}+\Delta L_i\gamma_{SL})\delta R_i+\delta E_{ki}$. And a stabilized system must obey that
	\begin{equation}\label{stabilized state}
		\frac{\delta E_i}{\delta R_i}=0
	\end{equation}
	A combination of equation \ref{internal energy} and equation \ref{stabilized state} leads to  
	\begin{equation}
		\cos\theta_{Li}=\cos\theta_0-\frac{1}{\gamma_{LG}}\left[ \frac{5}{2}ph+h\frac{p_0}{2}\exp( -\frac{L_{v,m}}{RT}) \right]-\frac{\gamma_{SG}}{\gamma_{LG}}
	\end{equation}
	where $p=p(R_i\cos\theta_{\Delta L_i},R_i\sin\theta_{\Delta L_i}),h=h(R_i\cos\theta_{\Delta L_i},R_i\sin\theta_{\Delta L_i})$, i.e., $p$ and $h$ are the pressure and thickness of soild-liquid interface boundary respectively. 
	
	Hence, we can get the mean contact angle with intergrating contact angle $\cos\theta_{Li}$ along the boundary. 
	\begin{equation}\label{mean contact angle}
		\cos\theta_L=\cos\theta_0-\frac{1}{L\gamma_{LG}}     \oint_{L}       \left[ \frac{5}{2}ph+h\frac{p_0}{2}\exp( -\frac{L_{v,m}}{RT}) \right]\mathrm{d}l-\frac{\gamma_{SG}}{\gamma_{LG}}
	\end{equation}
	where $\theta_0$ is the contact angle obeying the Young Equation, $L$ is the circumference of solid-liquid interface boundary, $L_{\mathrm{v,\:m}}$ is the latent heat of phase transition from liquid to gas phase. $T$ is the temperature of gas film. The element on the right side of equation \ref{mean contact angle} is the influence of gas film. The element on the middle of equation \ref{mean contact angle} is the influence of moving substrate. 
	
	We could conclude that the contact angle on moving substrate is 8$^{\circ}$ bigger than the one obeyed Young Equation in the room tempurature approximately. So the hydrophobicity will be reinforced on the moving substrate.  And the contact time\cite{11}\cite{12}, spreading factor\cite{12}, bouncing et.al. will change with the change of hydrophobicity between the droplet and substrate. Then we will elucidate them in detial. 
	\subsection{The analytical solution for Reynolds Equation}
	Firstly, we could get another equation which describes the gas film on the moving surface from Reynolds transport equation: 
	\begin{equation}
		\frac{\partial h}{\partial t}+\nabla\cdot\left( h\mathbf{u}\right) =0,\:\frac{\partial h}{\partial t}=0
	\end{equation}
	where $\mathbf{u}$ could be seen as the surface speed $U\mathbf{i}+V\mathbf{j}$ because of the Navier Boundary Conditions.
	
	Then, the problem is solving the differential equations:
	\begin{equation}\label{r1}
		\left\{\begin{matrix}
			h\frac{\partial^2p}{\partial x^2}+h\frac{\partial^2p}{\partial y^2}+3\left( \frac{\partial h}{\partial x}\frac{\partial p}{\partial x}+\frac{\partial h}{\partial y}\frac{\partial p}{\partial y}\right) =0
			\\U\frac{\partial h}{\partial x}+V\frac{\partial h}{\partial y}=0
			
		\end{matrix}\right.
	\end{equation}
	Then, we order that $h\left( x,y\right) =h_X\left( x\right) h_Y\left( y\right) $, $p\left( x,y\right) =p_X\left( x\right) p_Y\left( y\right) $. We could get equation \ref{r2} through bringing these to equation \ref{r1}.
	\begin{equation}\label{r2}
		\frac{1}{p_X^2}\frac{\mathrm{d}^2p_X}{\mathrm{d}x^2}+\frac{1}{p_Y^2}\frac{\mathrm{d}^2p_Y}{\mathrm{d}x^2}+\frac{3}{p_X^2p_Yh_X}\frac{\mathrm{d}h_X}{\mathrm{d}x}\frac{\mathrm{d}p_X}{\mathrm{d}x}+\frac{3}{p_Y^2p_Xh_Y}\frac{\mathrm{d}h_Y}{\mathrm{d}y}\frac{\mathrm{d}p_X}{\mathrm{d}y}=0
	\end{equation}
	Then, finding the derivative of equation \ref{r2} with respect to x and y in turn. We can get 
	\begin{equation}\label{r3}
		\frac{\mathrm{d}^2p_X}{\mathrm{d}x^2}-C\frac{\mathrm{d}p_X}{\mathrm{d}x}+C^\prime p_X^2=0
	\end{equation}
	\begin{equation}\label{r4}
		\frac{\mathrm{d}p_Y}{\mathrm{d}y}\frac{\mathrm{d}h_Y}{\mathrm{d}y}-\frac{Cp_Y^2h_Y}{3}=C^\prime \frac{p_Yh_Y}{3}
	\end{equation}
	$C,\:C^\prime $are constant. And we could esaily find the solution of equarion \ref{r1}.
	\begin{equation}
		h_X=C_{h_1}\exp\left( -\frac{\lambda}{U}x\right) ,\:h_Y=C_{h_2}\exp\left( \frac{\lambda}{V}y\right) 
	\end{equation}
	bring them to equation \ref{r4}, we can get 
	\begin{equation}
		\int \frac{\mathrm{d}p_Y}{C_2^\prime p_Y+C_2p_Y}=y
	\end{equation}
	So the $p_Y$ is 
	\begin{equation}
		p_Y=\frac{C_2}{-C_2^\prime +\exp\left( \frac{-y+C_2^{\prime\prime}}{C_2}\right) }
	\end{equation}
	Then, we solve the equation \ref{r3} with series method. Considering the series solution $p_X=\sum_{n=0}^{\infty}a_nx^n$. Then, we can get
	\begin{equation}
		a_{n+2}\left( n+2\right) \left( n+1\right) -Ca_{n+1}\left( n+1\right) +2C^\prime \left( a_0a_{n}+a_1+a_{n-1}+\cdots +a_{n/2}a_{n/2}\right) =0,\:\mathrm{n\:is\:an\:even.}
	\end{equation}
	\begin{equation}
		a_{n+2}\left( n+2\right) \left( n+1\right) -Ca_{n+1}\left( n+1\right) +2C^\prime \left( a_0a_{n}+a_1+a_{n-1}+\cdots +a_{\left( n-1\right) /2}a_{\left( n+1\right) /2}\right) =0,\:\mathrm{n\:is\:an\:odd.}
	\end{equation}
	And the radius of convergence $R$ obey that 
	\begin{equation}\label{r6}
		\lim_{n\to \infty}\left[ n+2+\frac{2C^\prime }{n+1}\left( a_0R^2+a_1R^3+\cdots +a_{n/2}R^{n/2+2}\right) \right]-CR=0,\:\mathrm{n\:is\:an\:even.}
	\end{equation}
	\begin{equation}\label{r7}
		\lim_{n\to \infty}\left[ n+2+\frac{2C^\prime }{n+1}\left( a_0R^2+a_1R^3+\cdots +a_{\left( n-1\right) /2}R^{\left( n+3\right) /2}\right) \right]-CR=0,\:\mathrm{n\:is\:an\:odd.}
	\end{equation}
	In addition, we can get equation \ref{r5} when $R=1$.
	\begin{equation}\label{r5}
		\begin{split}
			\lim_{n\to \infty}\left[ n+2+\frac{2C^\prime q\left( n/2+1\right) }{n+1}-C\right] \le&\lim_{n\to \infty}\left[ n+2+\frac{2C^\prime }{n+1}\left( a_0R^2+a_1R^3+\cdots +a_{n/2}R^{n/2+2}\right) \right]-CR\\
			&\le\lim_{n\to \infty}\left[ n+2+\frac{2C^\prime q^\prime\left( n/2+1\right) }{n+1}-C\right],\:\mathrm{n\:is\:an\:even.}\\
			\lim_{n\to \infty}\left[ n+2+\frac{2C^\prime w\left( \left( n+1\right) /2\right) }{n+1}-C\right] \le&\lim_{n\to \infty}\left[ n+2+\frac{2C^\prime }{n+1}\left( a_0R^2+a_1R^3+\cdots +a_{\left( n-1\right) /2}R^{\left( n+3\right) /2}\right) \right]-CR\\
			&\le\lim_{n\to \infty}\left[ n+2+\frac{2C^\prime w^\prime\left(\left( n+1\right) /2\right) }{n+1}-C\right],\:\mathrm{n\:is\:an\:odd.}\\
		\end{split}
	\end{equation}
	where $q=\min\left\lbrace a_0,\:a_1,\cdots,\:a_{n/2}\right\rbrace,\:q^\prime=\max\left\lbrace a_0,\:a_1,\cdots,\:a_{n/2}\right\rbrace  $ and $w=\min\left\lbrace a_0,\:a_1,\cdots,\:a_{\left( n-1\right) /2}\right\rbrace,\:w^\prime=\max\left\lbrace a_0,\:a_1,\cdots,\:a_{\left( n-1\right) /2}\right\rbrace  $. 
	
	If the equation \ref{r5} is right, the equation \ref{r6} and \ref{r7} would be wrong. So the $R$ is either $\infty$ or $1<R<\infty$. 
	
	So $p\left( x,y\right) $ is 
	\begin{equation}
		p=\left( a_0+a_1x+\frac{Ca_1-C^\prime a_0^2}{2}x^2+\cdots\right) \left( \frac{C_2}{-C_2^\prime +\exp\left( \frac{-y+C_2^{\prime\prime}}{C_2}\right) }\right) ,\:1<x\le \infty
	\end{equation}
	So $h\left( x,\:y\right) $ is
	\begin{equation}
		h\left( x,\:y\right) =h_X\cdot h_Y=C_{h}\exp\left( -\frac{\lambda}{U}x+ \frac{\lambda}{V}y\right)
	\end{equation}
	\section{Conclusion}
	In section 2, we find that how the moving substrate affect the hydrophobicity of droplet and we discuss the analytical solution for Reynolds Equation. Therefore, we would analytically get the contact angle on moving substrate. But we must have some boundary conditions such as $h\left( x,\:y\right)|_{\mathrm{droplet\:boundary}}=H\left( x,\:y\right) ,\:p\left( x,\:y\right)|_{\mathrm{droplet\:boundary}}=P\left( x,\:y\right) $ and so on to get the whole solution. In section 3, we find out some promotion and inhibition effects for bouncing question. Finally, we get a completely bouncing criteria $\mathrm{Cr}$ for droplet on moving substrate. Some phenomenons could be pridicted by using this criteria. In section 4, we research the effect of extra wind field for droplet. we find that extra wind field could change the final states of droplet. In addition, we get a completely bouncing criteria $\mathrm{Cr_{Wind}}$ for droplet on moving substrate with extra wind by the way that introduce the lift force potential. We also get the splashing criteria $\beta^2_{\mathrm{Wind}}$ using the R\&G Model. 
	
	But we don't analytically get a criteria because of the complex viscous dissipation. We just use a simple model to calculate it. We also don't consider the energy dissipation due to the roughness of substrate. It is so important that cound't be ignored in some substrate with big roughness. 
	\begin{acknowledgments}
	Thanks for Shangqian Sun, Hongwang Lu, Jingcheng Hao and Ying Ma 's support for this work.
	\end{acknowledgments}
	\bibliography{paper}
\end{document}